\documentclass[review]{elsarticle}

\usepackage{lineno,hyperref}
\usepackage{graphicx,amsmath,subfigure}
\usepackage{xcolor}
\modulolinenumbers[5]
\usepackage{graphicx}
\usepackage{subcaption}
\usepackage{epstopdf,epsfig}
\graphicspath{{figs/}}
\graphicspath{{figure/}}

\usepackage{verbatim}
\usepackage{newtxtext}
\usepackage{newtxmath}
\usepackage{natbib}
\usepackage{hyperref}
\hypersetup{
    colorlinks = true,
    urlcolor   = blue,
    citecolor  = black,
}

\newcommand{\RomanNumeralCaps}[1]
\linenumbers

\journal{XXX}

\begin{document}
\newcommand\blue{\textcolor[rgb]{0.00,0.00,1.00}} 
\newcommand\red{\textcolor[rgb]{1.00,0.00,0.00}}  
\begin{frontmatter}

\title{A symmetry-based theory for the mean velocity profiles of turbulent boundary layers subjected to pressure gradient and historical turbulence effects}


\author[addr1]{Tan-Tan Du}
\author[addr1]{Jun Chen}
\author[addr1]{Wei-Tao Bi}
\author[addr1]{Zhen-Su She\corref{mycorrespondingauthor}}
\address[addr1]{State Key Laboratory for Turbulence and Complex Systems and Department of Mechanics and Engineering Science, College of Engineering, Peking University, Beijing 100871, P.R. China}
\cortext[mycorrespondingauthor]{Corresponding author}
\ead{she@pku.edu.cn}

%
%

\begin{abstract}
The pressure gradient (PG) has a significant influence on the mean-flow properties of turbulent boundary layers (TBLs). Conventional analytical studies on PG TBLs are conducted separately for the viscous sublayer and overlap layer only, with the remaining large portion of the boundary layers less described theoretically. Here a symmetry approach is proposed for modeling whole mean velocity profiles of TBLs with both PG and historical turbulence effects. First, a modified defect law is constructed for the total stress profile to capture the Reynolds stress overshoot owing to the PG and historical turbulence effects. Second, the multi-layered power-law formulation of the stress length function in the structural ensemble dynamics theory of the canonical zero-PG TBL (J. Fluid Mech., 2017, Vol. 827, pp. 322-35) is extended to describe the PG TBL. Comparing with that of the zero-PG TBL, the stress length of the PG TBL possesses a variable (in magnitude and extension) plateau in the defect layer, which are characterized by three parameters: the buffer layer thickness, the (nominal) K\'{a}rm\'{a}n constant, and the defect-law exponent. In the case of intense turbulence from upstream flow, a newly-identified ``bulk turbulent layer" replaces the conventional buffer layer and overlap layer. With the above formulations, the entire mean velocity profile is predicted analytically, and validated to accurately describe the published direct numerical simulation data on a two-dimensional separation bubble. The study provides a novel parameterization for PG TBLs with high accuracy and sound physics, and paves a way for quantifying complicated boundary-layer flows.
\end{abstract}

\begin{keyword}
Turbulent boundary layer \sep Pressure gradient \sep Turbulence theory \sep Symmetry
\end{keyword}

\end{frontmatter}


\section{Introduction}
\label{sec:Introduction}
Turbulent boundary layers (TBLs) on the surfaces of flying, ground, and underwater vehicles, wind and turbo-engine blades, etc., are generally subjected to non-zero pressure gradient (PG). Compared with the canonical zero-pressure-gradient (ZPG) TBL, TBLs with adverse and favorable pressure gradients (APGs/FPGs) are more intriguing and less understood, because of the non-locality nature of pressure, the expanded parametric space (e.g. surface curvature), and the relevant complex flow phenomena such as separation.

A typical feature of the PG TBLs is the classical law of the wall and defect law cease to be valid. To retrieve a similar solution people have studied so-called equilibrium TBLs \citep{clauser1954turbulent,bradshaw1967turbulence}, which possess a similar velocity-deficit profile in the outer region. Equilibrium TBLs occur when the free-stream velocity exhibits a power-law variation with the streamwise coordinate \citep{townsend1961equilibrium}, which seldom happens in engineering flows. To find a solution for general PG TBLs, researchers have attempted to identify the relevant velocity and length scales. In addition to the friction velocity $u_\tau$, a PG-based velocity scale has been introduced: ${u_p} \equiv \sqrt[3]{( {{\nu }/{\rho })({{\partial p }}/{{\partial x}}})}$, where $\nu$ is kinetic viscosity, $\rho$ is density, $p$ is the mean pressure, and $x$ is the streamwise coordinate. Based on ${u_p}$, \cite{stratford1959prediction} obtained a square root law for the velocity profile with zero wall shear. \cite{townsend1961equilibrium} extended the theory to PG TBLs with positive wall shear and derived an expression for the mean velocity profile that includes both square root and logarithmic parts. \cite{skote2002direct} revisited Townsend's work through an argument of overlap of inner and outer flows. \cite{ye2017near} further extended the work to compressible flows, specifically oblique shock wave/turbulent boundary layer interactions. Note that these analytical solutions for the mean velocity were derived separately for the viscous sublayer and overlap region only. For the outer flow, formulations have to be constructed empirically. The most influential work is the law of the wake by \cite{coles1956law}. \cite{perry2002streamwise} studied the parameter $\mathnormal\Pi$ in Coles's law in general non-equilibrium TBLs. Numerous studies were also conducted on identifying the relevant scales for acquiring an outer-flow scaling \citep{schofield1972turbulent,zagarola1998mean,maciel2006self,gungor2016scaling,devenport2022equilibrium}. \cite{castillo2001similarity} extended the concept of equilibrium TBL and introduced a PG parameter to characterize the velocity profile in the outer flow. Very recently \cite{wei_knopp_2023} used the maximum Reynolds shear stress location to determine the proper length scale, and proposed an outer scaling that collapses well the experimental and numerical data on APG TBLs. Whereas these studies have empirically proposed similar velocity profiles for the outer flow, there seldom have been analytical expressions to describe the profiles. A theoretical framework beyond the conventional approaches is needed in order to construct an analytical solution of the entire mean velocity profile of the PG TBLs.

Recently,  \cite{she2010new,she2017quantifying} proposed a structural ensemble dynamics (SED) theory for the canonical ZPG TBL. In SED, the constraint imposed by the wall on the turbulence is expressed through a stress length (SL) function that exhibits multilayered dilation symmetry with respect to the wall distance, forming a so-called multilayer structure (MLS) that arises due to the conversion of balance mechanisms away from the wall in the transport equation of turbulent fluctuations \citep{chen2018quantifying}. By introducing a simple dilation-symmetry-breaking ansatz to represent the transition of dilation symmetry between adjacent layers, the SED provides excellent analytical descriptions of experimentally and numerically observed mean profiles of canonical wall turbulence (i.e., channel, circular pipe, and ZPG TBL). As for the non-canonical TBLs with various effects such as compressibility, roughness, and PG, the SED assumes that the dilation symmetry is preserved because of the dominant role of the wall constraint, but the MLS is limitedly perturbed by the effects. Thus, with variable MLS parameters to quantify the perturbed MLS, the SED solution can be extended to describe non-canonical TBLs, such as rough pipe TBLs \citep{she2012multi}, transitional boundary layers \citep{xiao2019symmetry}, and atmospheric TBLs \citep{ji2021analytic}. In this paper, the symmetry approach of the SED is extended to describe the PG TBLs.

Here, a modified defect law is constructed for the total stress profile to quantify the PG and historical turbulence effects. The variation of the MLS with PG is characterized by three MLS parameters: the buffer layer thickness, the (nominal) K\'{a}rm\'{a}n constant, and the defect-law exponent. The influence of the upstream turbulence on the MLS is characterzed with a new layer called ``bulk turbulent layer". With the symmetry-based formulations of the total stress and SL function, the entire mean velocity profiles are predicted and validated with the recent direct numerical simulation (DNS) data on a two-dimensional (2D) incompressible separation bubble flow. The study presents a novel parameterization for the mean-flow profiles of general PG TBLs, and paves a way to quantify the PG effects with both great accuracy and sound physics.

\section{Theory}
\subsection{The conventional analysis}
The streamwise mean momentum equation of a 2D incompressible TBL reads
 \begin{equation}
  u\frac{\partial u}{\partial x}+v\frac{\partial u}{\partial y}  = -\frac{1}{\rho} \frac{\partial p}{\partial x}+\nu \left(\frac{\partial^2 u}{\partial x^2}+\frac{\partial^2 u}{\partial y^2}\right)-\frac{\partial \overline{u'u'}}{\partial x}-\frac{\partial \overline{u'v'}}{\partial y},
  \label{eq:x_momentum}
\end{equation}
where $x$ and $y$ denote the streamwise and wall-normal coordinates, $u$ and $v$ represent the streamwise and wall-normal mean velocity components, respectively, with $u'$ and $v'$ representing the corresponding velocity fluctuations, and the overline indicates Reynolds averaging. The wall-normal integration of (\ref{eq:x_momentum}) leads to
 \begin{equation}
  \tau\equiv\nu \frac{\partial u}{\partial y}-\overline{u'v'}=\nu \frac{\partial u}{\partial y}\Big|_w +\int_{0}^{y}{\left(u\frac{\partial u}{\partial x}+v\frac{\partial u}{\partial y}+\frac{1}{\rho} \frac{\partial p}{\partial x}-\nu \frac{\partial^2 u}{\partial x^2}+\frac{\partial \overline{u'u'}}{\partial x}\right)dy},
  \label{eq:tau}
\end{equation}
where subscript $w$ denotes the variables on the wall, and $\tau$ is the total stress. From the integral on the right hand side of (\ref{eq:tau}), the $\tau$ profile is crucially determined by the convection and the streamwise PG.

By modeling the total stress and Reynolds stress, the streamwise mean velocity profile can be predicted through (\ref{eq:tau}). Conventional approaches are conducted in two regions: the viscous sublayer and the overlap layer. Within the viscous sublayer, the Reynolds stress is negligible and the total stress can be approximated with
\begin{equation}
  \tau^+=1 +P_w^+ y^+,
  \label{eq:tau_nearwall}
\end{equation}
where superscript plus denotes the wall-unit normalization, and $P_w^+ = [\nu/(\rho u_\tau^3)]({\partial p}/{\partial x})_w$. Subsequently, the streamwise mean velocity in the viscous sublayer can be derived from (\ref{eq:tau}) and (\ref{eq:tau_nearwall}) to give $u^+=y^+ +P_w^+ {y^+}^2/2$. Within the overlap region, the Reynolds stress dominates and can be modeled with Prandtl's mixing length hypothesis \citep{prandtl1925ausgebildete}: $-\overline{u'v'}=\ell_m^2\left({\partial u}/{\partial y}\right)^2$, where $\ell_m=\kappa y$ and $\kappa$ represents the K\'{a}rm\'{a}n constant. With (\ref{eq:tau_nearwall}) and the mixing length closure, $u$ in the overlap region can be solved; the expression is complicated, but one may refer to \cite{townsend1961equilibrium} and \cite{skote2002direct} for the details. For the other boundary-layer regions, the conventional analysis cannot be conducted.

Here we present a novel symmetry-based approach. The symmetry approach aims to describe the dilation symmetry (i.e. power law with respect to the wall distance) exhibited by TBL eddies under the wall constraint. In the SED theory, three dilation symmetries have been identified for wall turbulence: a normal power law, a defect power law, and a dilation-symmetry-breaking ansatz. A combination of the dilation symmetries makes up analytical expressions for the whole total stress and mixing length (recalled SL in the SED) profiles, yielding new prediction of the entire mean velocity profile for both the APG and FPG TBLs.

\subsection {The modified defect law of the total stress}\label{sec:tau}
Let us first investigate the dilation symmetry of the total stress. For the canonical wall turbulence, it has been proposed that $\tau$ obeys the following defect power law:
 \begin{equation}
  \tau ^ +  = 1-\left(\frac{y^+}{\delta^+}\right)^{\alpha},
  \label{eq:tau_ZPG}
\end{equation}
where $\delta$ is the boundary-layer thickness ($\delta=\delta_{99}$ for a ZPG TBL, $\delta=R$ for a channel or pipe, where $R$ is the half-width of channel or radius of pipe), and $\alpha$ is the defect-power-law exponent. $\alpha$ is exactly unity for channel and pipe flows, and was estimated to be $3/2$ for most of the canonical ZPG TBL by \cite{chen2016analytic} through exploring experimental and DNS data (note that $\alpha=3$ in the viscous sublayer).

In case of a TBL with a non-zero PG, (\ref{eq:tau_ZPG}) is no longer applicable and must be revised to account for the PG effect. Here, we construct a modified defect power law for the total stress profile of PG TBL as follows:
 \begin{equation}
  \tau_{PG} ^ +(P_0^+,y^+)  = 1 -\left(\frac{y^+}{\delta^+}\right)^{3/2}{{ + P}}_0^{{ + }}\left[ \left({\frac{y^+}{y_P^ + }}\right) {{\left( {1 + {{\left( {\frac{y^+}{y_P^ + }} \right)}^2}} \right)}^{-1/2} - \left(\frac{y^+}{\delta^+}\right)^{3/2}} \right],
  \label{eq:tau_PG}
\end{equation}
where $y_P^+=P_0^+/P_w^+$, and $P_0^+$ denotes a characteristic PG-related stress to be determined with experimental and numerical data. Although not being a key issue here, the definition of $\delta$ is problematic for PG TBLs \citep{vinuesa2016determining}. Here, we define $\delta$ as the thickness where $y\partial u/\partial y$ reduces by half from its second peak when $y$ approaches the potential flow \citep{DuTT2023}. This definition has the advantage of being approximately $\delta_{99}$ in ZPG TBLs. The derivation of (\ref{eq:tau_PG}) is explained as follows. Firstly, (\ref{eq:tau_PG}) satisfies the asymptotic condition. For the canonical ZPG TBL where $P_0^+=0$, (\ref{eq:tau_PG}) reduces to (\ref{eq:tau_ZPG}). Near the wall where $y^+\ll y_P^+$, (\ref{eq:tau_PG}) reduces to (\ref{eq:tau_nearwall}). In the outer flow where $y^+\gg y_P^+$, (\ref{eq:tau_PG}) reduces to
 \begin{equation}
  \tau ^ +  = \left(1+P_0^+\right) \left[1
  -\left(\frac{y^+}{\delta^+}\right)^{3/2}\right].
  \label{eq:tau_bulk}
\end{equation}
(\ref{eq:tau_bulk}) assumes that the Reynolds stress profile in the outer region of PG TBL exhibits dilation symmetry similar to that of the canonical ZPG TBL. Compared to the maximum value of unity for the Reynolds stress profile in the ZPG TBL, $P_0^+$ then represents a measure of the Reynolds stress overshoot in the APG TBL, as well as a measure of the Reynolds stress deficit in the FPG TBL.

The second aspect of the derivation of (\ref{eq:tau_PG}) involves connecting the near-wall and outer-flow expressions of $\tau^+$ (i.e. (\ref{eq:tau_nearwall}) and (\ref{eq:tau_bulk})) through the dilation-symmetry-breaking ansatz. In (\ref{eq:tau_PG}) the ansatz reads: $[ {1 + ( y^+/y_P^ + )}^2 ]^{-1/2}$, which describes a continuous transition, occurring at the thickness $y_P^+$, from a power law of ${y^+}^1$ to ${y^+}^{0}$. Such an ansatz has been validated in various complex turbulent systems as a universal description of the dilation-symmetry-breaking process, and here it is applied to model the total stress profile.

(\ref{eq:tau_PG}) has to be further extended to describe the reattached FPG TBLs, where intense turbulent fluctuations from the upstream free shear layer may fill the boundary layer and significantly increase the Reynolds stress. In our previous study on the reattached boundary layers of compression ramps \citep{hu2017beta}, this excess Reynolds stress (denoted by $W^+$) was modeled with the so-called \textit{Beta}-distribution, which is a dual-power-law that can be written as,
 \begin{equation}
  W ^ + (W_{max}^+,y^+)=\frac{W_{max}^+}{0.0462}\left(\frac{y^+}{\delta^+}\right)^{1.7}\left(1-\frac{y^+}{\delta^+}\right)^3.
  \label{eq:tau_ex_Restress}
\end{equation}
In (\ref{eq:tau_ex_Restress}), $W_{max}^+$ represents the maximum excess Reynolds stress, 0.00462 is used for normalization such that the peak of the dual-power-law profile is $W_{max}^+$, and the power exponents 1.7 and 3 are determined from the compression ramp flows \citep[see][]{hu2017beta} and employed to describe the reattached TBLs in the present study. (\ref{eq:tau_ex_Restress}) indicates that the peak excess Reynolds stress is located constantly at about $0.362\delta$ during the relaxation process of the intense turbulence. Consequently, the total stress profile of the reattached TBL can be modeled by the sum of (\ref{eq:tau_PG}) and (\ref{eq:tau_ex_Restress}) as
 \begin{equation}
  \tau_{PG-Turb} ^ + (P_0^+,W_{max}^+,y^+)=\tau_{PG}^++W ^ + .
  \label{eq:tau_FPG}
\end{equation}
We found that a large number of FPG TBLs with historical turbulence effect can be described with (\ref{eq:tau_FPG}).

\subsection {A multilayered power law of the stress length}\label{sec:SL}
In order to predict the mean velocity profile, the Reynolds stress in (\ref{eq:tau}) can be modeled using the conventional mixing length argument. In the outer portion of a TBL (especially APG TBL) the eddies are less affected by the wall constraint and their size tends to depend on $\delta$ only. Therefore, the SL, which characterizes the eddy size according to the SED theory, is assumed to be a fraction of $\delta$ near the boundary layer edge: $\ell_m^+=\lambda \delta^+$, where $\lambda$ is a proportionality coefficient (which is indeed the turbulence model of \cite{escudier1966entrainment}, where $\lambda$ was taken to be $0.09$). This formulation can be extended to the entire outer flow by invoking a defect power law to derive \citep{she2017quantifying}:
\begin{equation}
  \ell_{out} ^ + = \lambda \delta^+ (1-r^n),
  \label{eq:ell_out}
\end{equation}
where $\ell_{out} ^ +$ represents the SL in the outer flow, $r = 1-y^+/\delta^+$ denotes the outer coordinate with origin at $\delta$ and unity at the wall, and $n$ is the defect-law exponent. The larger $n$ is, the wider the constant-SL region, and the stronger the APG. Note that as the overlap layer is approached (i.e. $r\approx1$),  (\ref{eq:ell_out}) reduces to the Prandtl's mixing length hypothesis with $\kappa=n\lambda$. For the canonical ZPG TBL, the SED theory has determined that $n=4$ and $\kappa=0.45$ \citep{she2017quantifying}, therefore $\lambda=0.1125$. For PG TBLs, the variations of $n$ and $\lambda$ (or $\kappa$) should be examined by exploring experimental and numerical data.

Regarding the SL profile in the inner region (denoted by $\ell_{in}^+$), a commonly-used expression employs the van Driest damping function to yield: $\ell_{in}^+=\kappa y^+ [(1-\mathrm{exp}(-y^+/A^+)]$, where $A^+=26$ \citep{van1956turbulent}. In contrast, the SED has proposed the following formula for $\ell_{in}^+$ by expressing the power laws and power-law transitions in the viscous sublayer, buffer layer and logarithmic layer \citep{she2017quantifying}:
\begin{equation}
  \ell_{in}^+=\frac{9.7^2\kappa}{y_b^+}\left(\frac{y^+}{9.7}\right)^{3/2}\left[1+\left(\frac{y^+}{9.7}\right)^{4}\right]^{1/8}
\left[1+\left(\frac{y^+}{y_{b}^+}\right)^{4}\right]^{-1/4},
  \label{eq:ell_in}
\end{equation}
where $9.7$ is the dimensionless sublayer thickness determined by the SED for the canonical wall turbulence and assumed invariant here for PG TBLs, and $y_{b}^+$ denotes the buffer layer thickness. According to the SED theory, $y_{b}^+$ is about $41$ in the canonical ZPG TBLs. In PG TBLs $y_b^+$ should be considered an empirical parameter. Then, the entire profile of the SL function can be constructed by combining (\ref{eq:ell_out}) and (\ref{eq:ell_in}) as follows,
\begin{equation}
  \ell_{m-PG}^+(n,\kappa,y_b^+,y^+)=\ell_{in}^+\ell_{out}^+/(\kappa y^+).
  \label{eq:ell}
\end{equation}
(\ref{eq:ell}) with the three PG-dependent parameters describes the MLS of a PG TBL. Note that \cite{subrahmanyam2022universal} recently proposed a different construction for $\ell_m^+$. Their formulation employs the $\ell_{in}^+$ of van Driest (with variable $A$ and an additional power-exponent in the exponential function to describe the PG effects), and uses the dilation-symmetry-breaking ansatz (instead of a defect law) for the $\ell_{out}^+$ that reads $\ell_{out}^+=\kappa y^+\left[1+\left(y^+/{(b\delta^+)}\right)^n\right]^{-1/n}$, where $\kappa$, $n$, and $b$ are three empirical parameters. With a different modeling for the total stress, \cite{subrahmanyam2022universal} claimed a unified expression for the mean velocity profile of wall turbulence including APG TBL. The differences between their approach and ours are that, the current study is consistently based on describing the dilation symmetry of the flows, posesses less empirical parameters with clear physical meaning, and can be extended to more complicated cases as discussed in the following.

As mentioned, the reattached FPG TBL is additionally subjected to intense turbulence coming from the separated shear layer. Before being fully dissipated through a long relaxation process, the turbulence occupies a large portion of the boundary layer and substantially alters the MLS, such that (\ref{eq:ell}) should be revised.

As is well-known, in a canonical ZPG TBL the most violent turbulent fluctuations occur in the buffer layer \citep{jimenez2018coherent}. Therefore, it is reasonable to guess that the intense turbulence from the upstream free shear layer stands as a new ``buffer layer" with greatly increased thickness. We rename this new ``buffer layer" as ``bulk turbulent layer", and its thickness as $y_{t}^+$. Because $y_{t}^+$ is elevated up to the outer flow, the conventional overlap layer disappears and is replaced by the bulk turbulent layer. Furthermore, because the bulk turbulent layer is to some extent similar to the free-shear-layer turbulence, the SL near $\delta$ is smaller than that at the centre of the bulk turbulent layer, leading to a unique decreasing profile with increasing $y$ beneath $\delta$. To describe the above features, we construct the MLS of the reattached TBL as follows:
\begin{equation}
  \ell_{m-PG-Turb}^+(\kappa,y_t^+,y^+)=\frac{9.7^{2.5}\kappa\delta^+}{{y_t^+}^{2.5}}\left(\frac{y^+}{9.7}\right)^{3/2}\left[1+\left(\frac{y^+}{9.7}\right)^{4}\right]^{1/4}
\left[1+\left(\frac{y^+}{y_{t}^+}\right)^{4}\right]^{-5/8}\frac{(1-r^4)}{4(1-r)},
  \label{eq:ell_FPG}
\end{equation}
which contains only two empirical parameters: $y_t^+$ and a nominal $\kappa$. Note that the bulk turbulent layer has a power exponent of 2.5 that is slightly larger than that of the buffer layer (which is 2), depicting the stronger turbulence in the bulk of the reattached TBL. Also note that $n$ in (\ref{eq:ell_out}) is set 4, which is adequate for describing most FPG TBLs before approaching relaminization.

Combining (\ref{eq:tau}), (\ref{eq:tau_PG}), and (\ref{eq:ell}) (or (\ref{eq:tau}), (\ref{eq:tau_FPG}), and (\ref{eq:ell_FPG}) for reattached TBLs), the streamwise mean velocity profile of a PG TBL can be calculated to be
 \begin{equation}
  u^+(y^+) =  \int_{0}^{y^+} {\frac{{ - 1 + \sqrt {4{\tau ^{\rm{ + }}}{{ {\ell_{m}^+}}^2}+ 1} }}{{2{{ {\ell_{m}^+}}^2}}}\mathrm{d} {y^{\rm{ + }}}}.
  \label{eq:u_model}
\end{equation}

\section{Validation}
The theory has been validated with dozens of experimental and numrical data available in the literature, for various flows including planar PG TBLs, airfoil flow, and TBL over Gaussian bump. Here we present the validation with the 2D separation bubble flow simulated by Abe \cite{abe2019direct}.

In Abe's DNS the suction and blowing are imposed at the upper boundary to induce a 2D separation bubble on a flat plate. The inlet TBL at $x=0$ has a momentum Reynolds number of 900. The TBL separates at about $x=120\theta_0$ ($\theta_0$ is the inlet momentum thickness) in front of which the TBL is subjected to APG, and reattaches at about $x=250\theta_0$ after which the TBL is subjected to FPG and strong turbulence from the separated shear layer.

\begin{figure}
    \centering
    \includegraphics[width=0.5\linewidth]{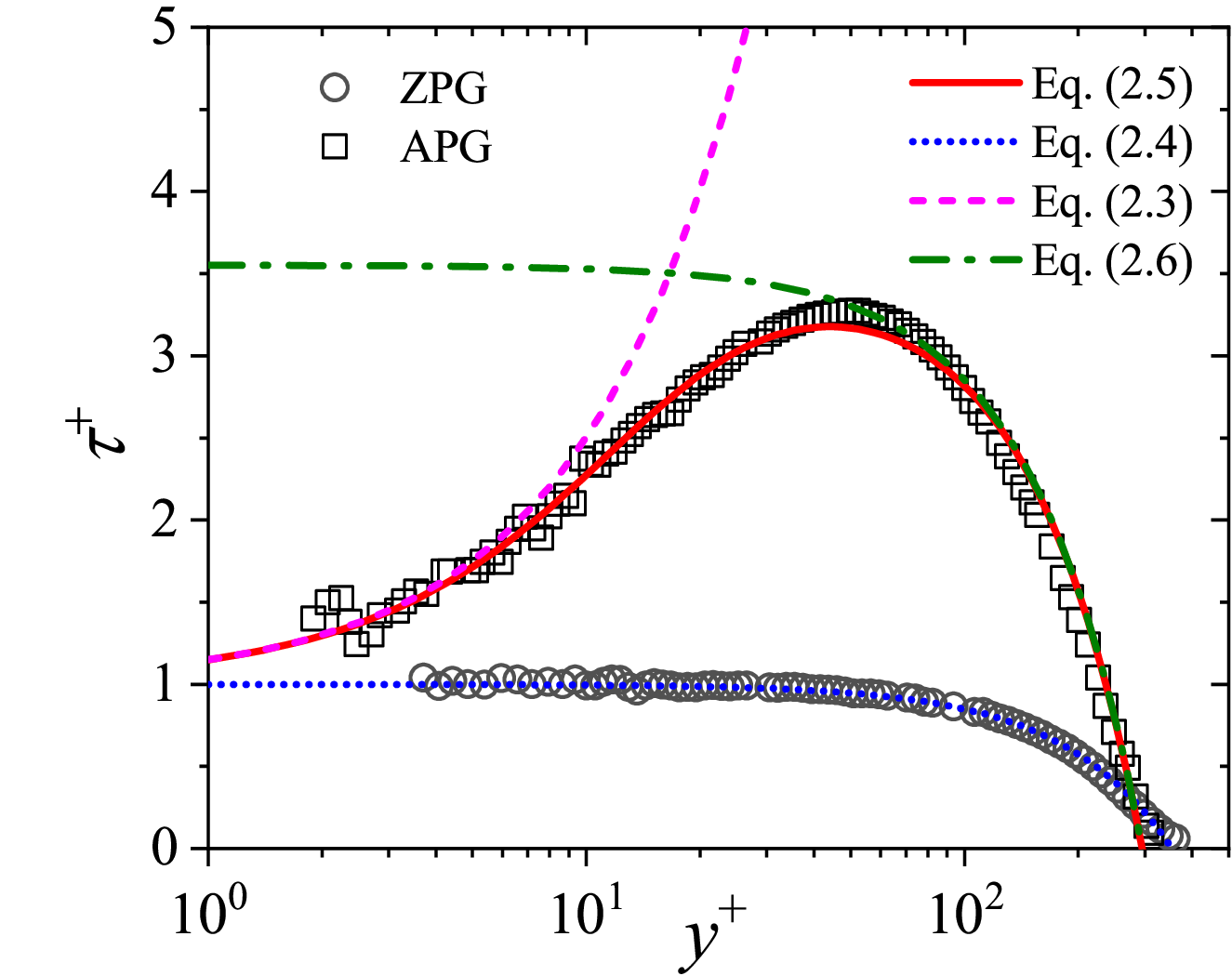}
    \includegraphics[width=0.5\linewidth]{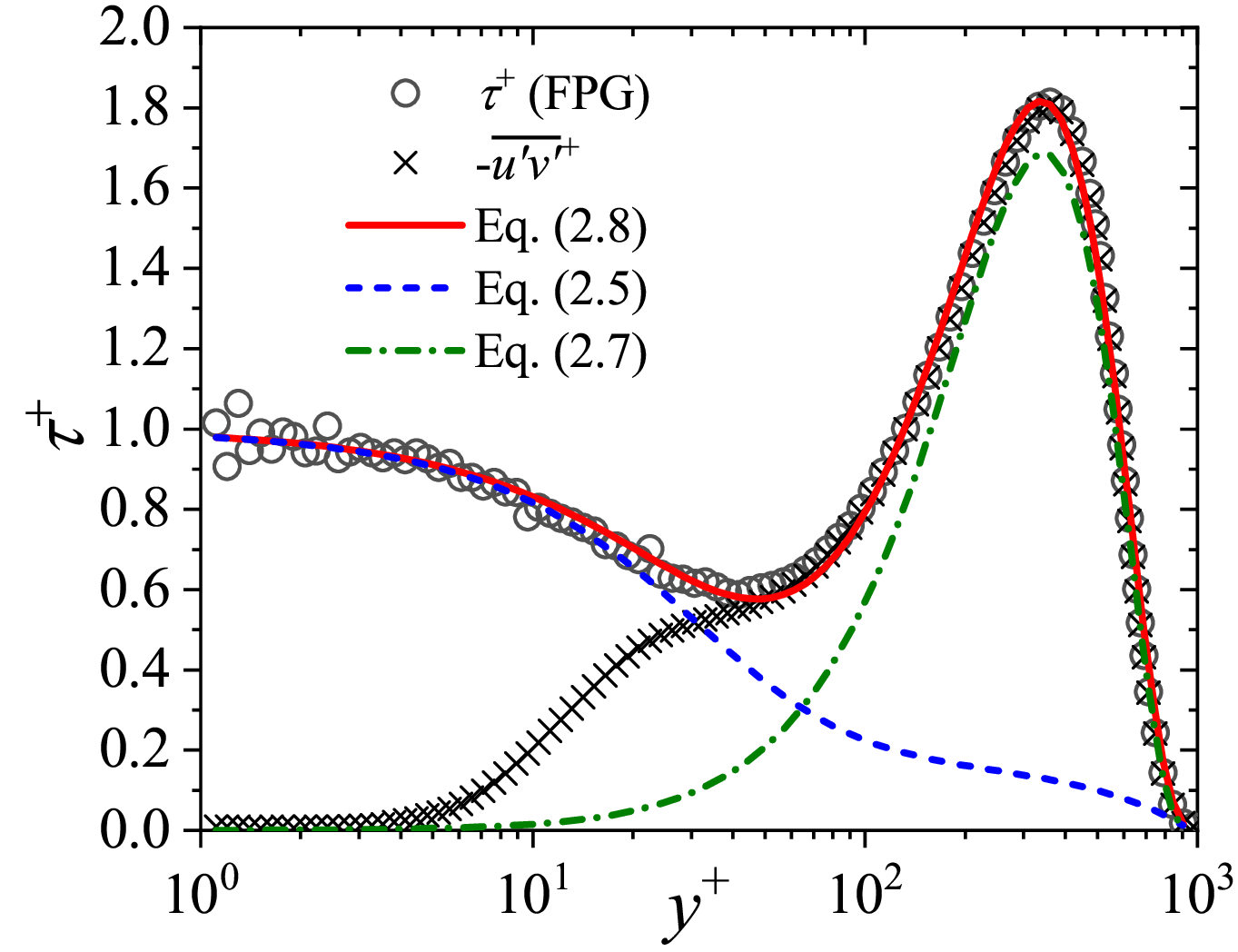}
  \caption{The total stress profiles of ($a$) the TBLs in front of a 2D separation bubble, and ($b$) the reattached TBL after the bubble. Symbols denote the DNS data digitized from \cite{abe2019direct} for the $Re_{\theta0}=900$ case. In ($a$), open circles denote the ZPG TBL at $x=0$ and open squares denote the APG TBL at $x=75\theta_0$. In ($b$), open circles and crosses denote the FPG TBL at $x=350\theta_0$.}
  \label{fig:tau_TBLs}
\end{figure}

Figure \ref{fig:tau_TBLs} displays the construction of the total stress model visualized with Abe's data. In figure \ref{fig:tau_TBLs}($a$) (\ref{eq:tau_ZPG}) and (\ref{eq:tau_PG}) rather accurately describe the $\tau$ profiles of ZPG and APG TBLs, respectively. The APG leads to a Reynolds stress overshoot in the bulk of the boundary layer, which is characterized by $P_0^+$, as indicated by the dash-dotted line of equation (\ref{eq:tau_bulk}). Therefore $P_0^+$ stands as a virtual wall shear stress owing to the APG. One finds that (\ref{eq:tau_PG}) smoothly connects the inner solution (\ref{eq:tau_nearwall}) and the outer solution (\ref{eq:tau_bulk}), which is achieved by the dilation-symmetry-breaking ansatz in (\ref{eq:tau_PG}). In FPG TBL, however, a negative $P_0^+$ results in a Reynolds stress deficit according to (\ref{eq:tau_PG}), as shown by the dashed line in figure \ref{fig:tau_TBLs}($b$). It is clear that (\ref{eq:tau_PG}) is not applicable to the reattached FPG TBL in Abe's DNS, owing to the intense turbulence in the bulk turbulent layer. As is shown, the turbulence leads to an excess Reynolds stress which is captured by the dual-power law of (\ref{eq:tau_ex_Restress}). With only two parameters, $P_0^+$ and $W_{max}^+$, the $\tau$ profile of the reattached FPG TBL is accurately described by (\ref{eq:tau_FPG}).

\begin{figure}
    \centering
    \includegraphics[width=0.5\linewidth]{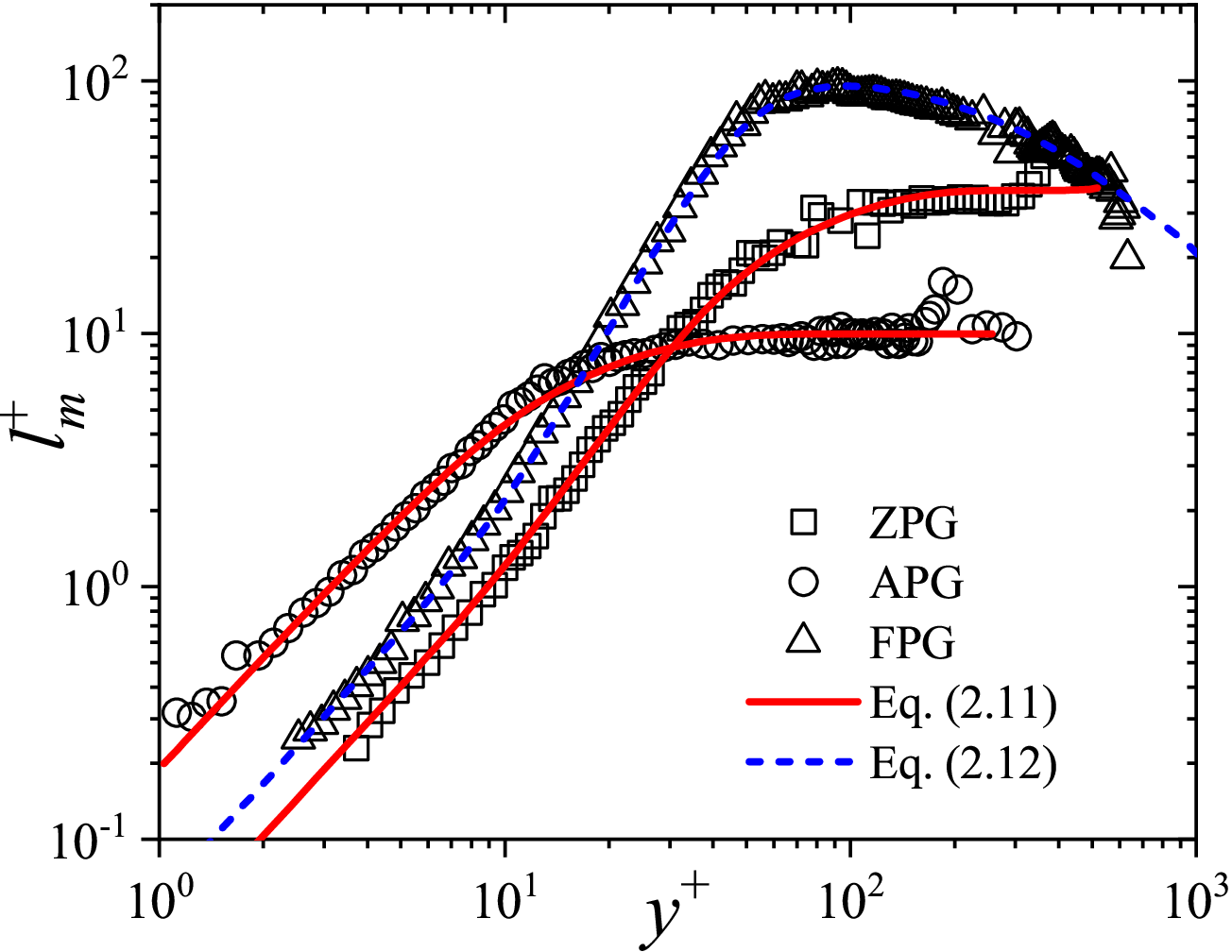}
  \caption{Validation of the stress length models with Abe's DNS. Squares denote the ZPG TBL at $x=0$, circles denote the APG TBL at $x=100\theta_0$, and triangles denote the reattached FPG TBL at $x=275\theta_0$.}
  \label{fig:stress-length}
\end{figure}

Then the SL models are validated in figure \ref{fig:stress-length} with Abe's DNS. The APG tends to relax the wall constraint and elevate the eddies \citep{maciel2017coherent}, and in the extreme conditions, the eddies become entirely unattached to wall such that the boundary layer separates. Therefore, comparing with those of the ZPG TBL, the buffer layer of the APG TBL becomes thinner, the outer flow with nearly constant SL becomes thicker, and the outer-flow eddy size relative to $\delta$ becomes smaller (because of the rapid increase of $\delta$ owing to the APG). These features show up in the SL profile of the APG TBL in figure \ref{fig:stress-length}, and are accurately described by (\ref{eq:ell}) with a smaller $y_b^+$, a larger $n$, and a smaller $\lambda$ than the corresponding ZPG ones. As to the reattached FPG TBL, a thick bulk turbulent layer occupies a crucial part of the boundary layer, characterized by a rapidly increasing SL and then a decreasing SL above the layer. As shown in figure \ref{fig:stress-length} (\ref{eq:ell_FPG}) rather accurately describes the SL profile of the reattached FPG TBL.

\begin{figure}
    \includegraphics[width=0.5\linewidth]{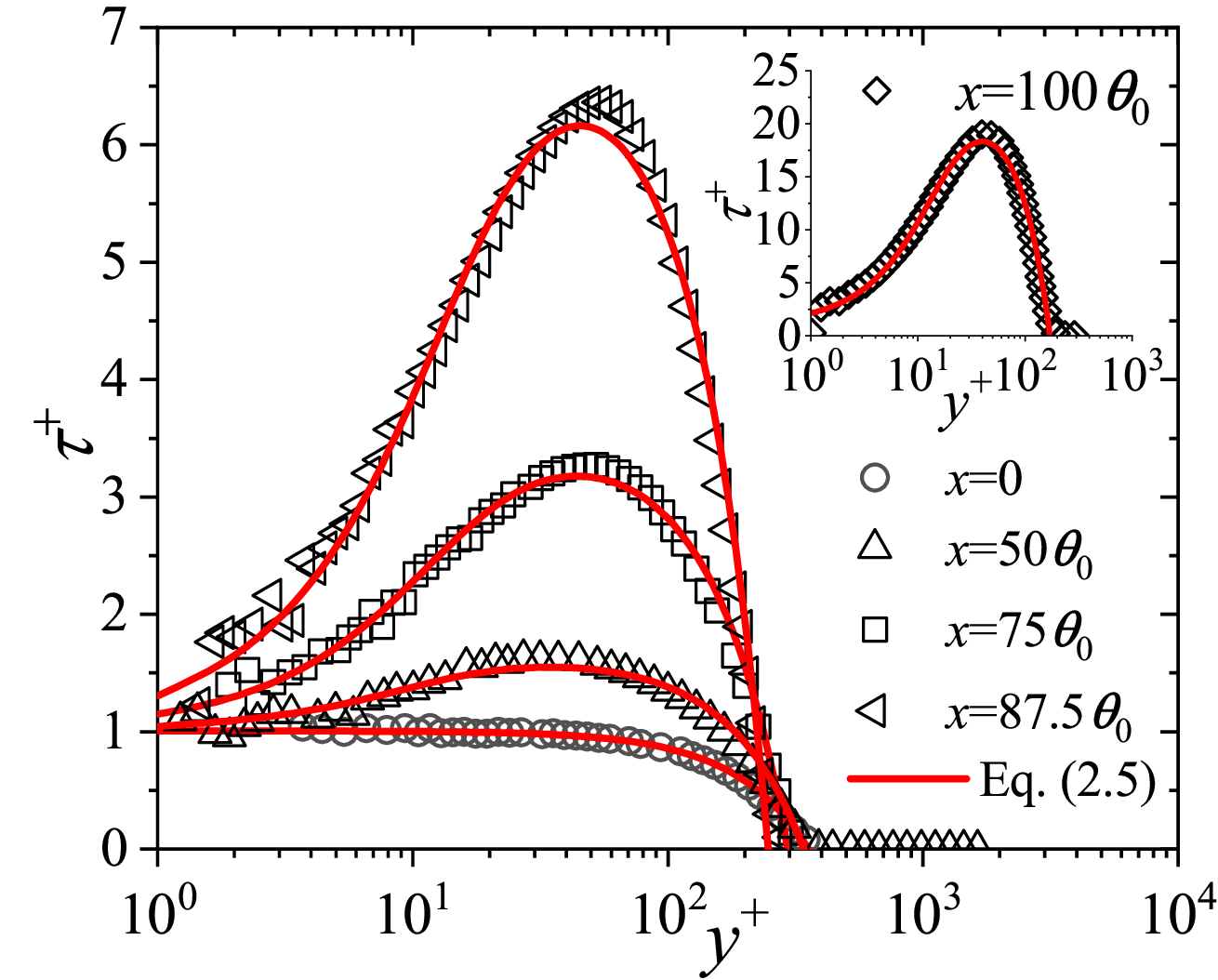}
    \includegraphics[width=0.5\linewidth]{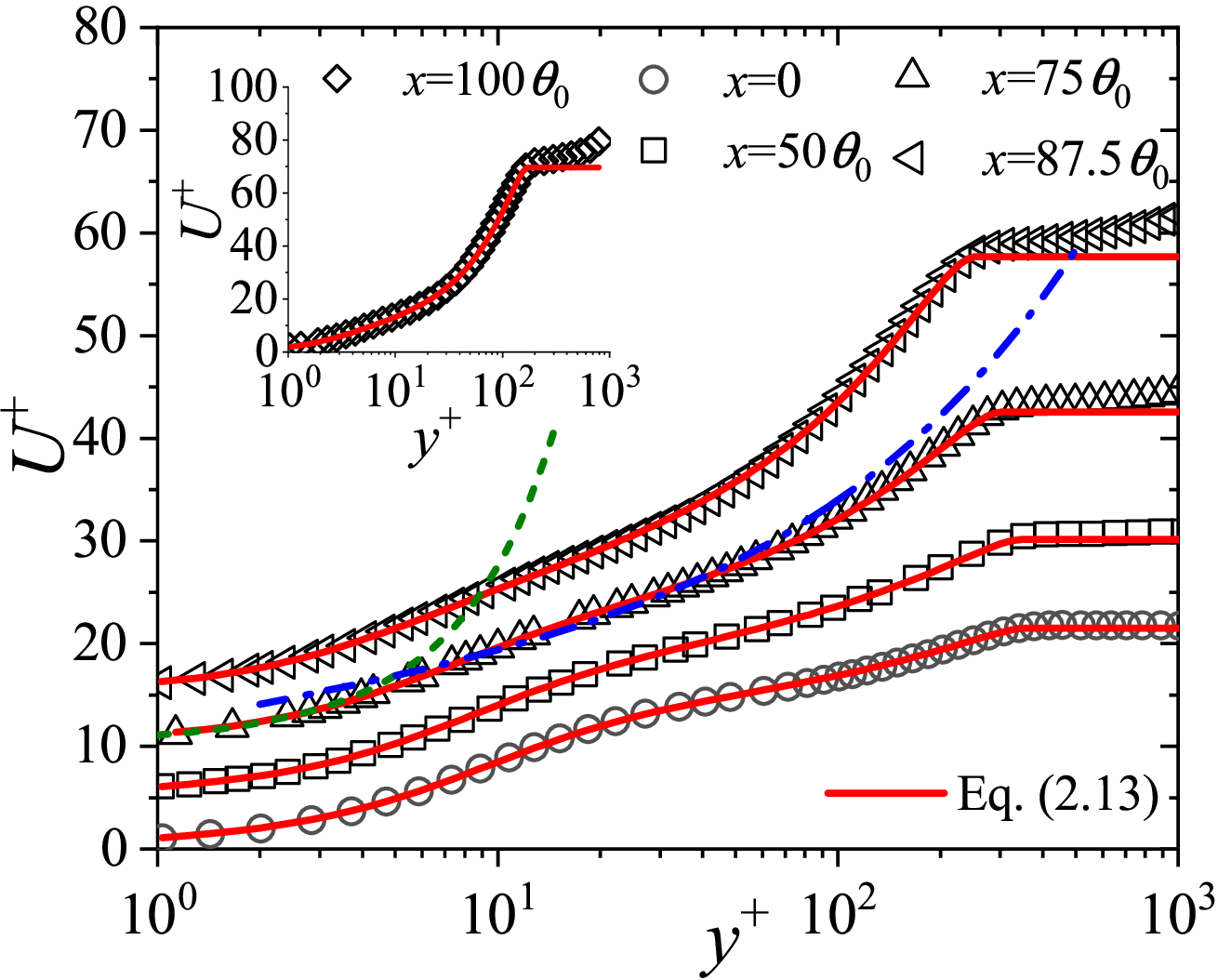}
    \includegraphics[width=0.5\linewidth]{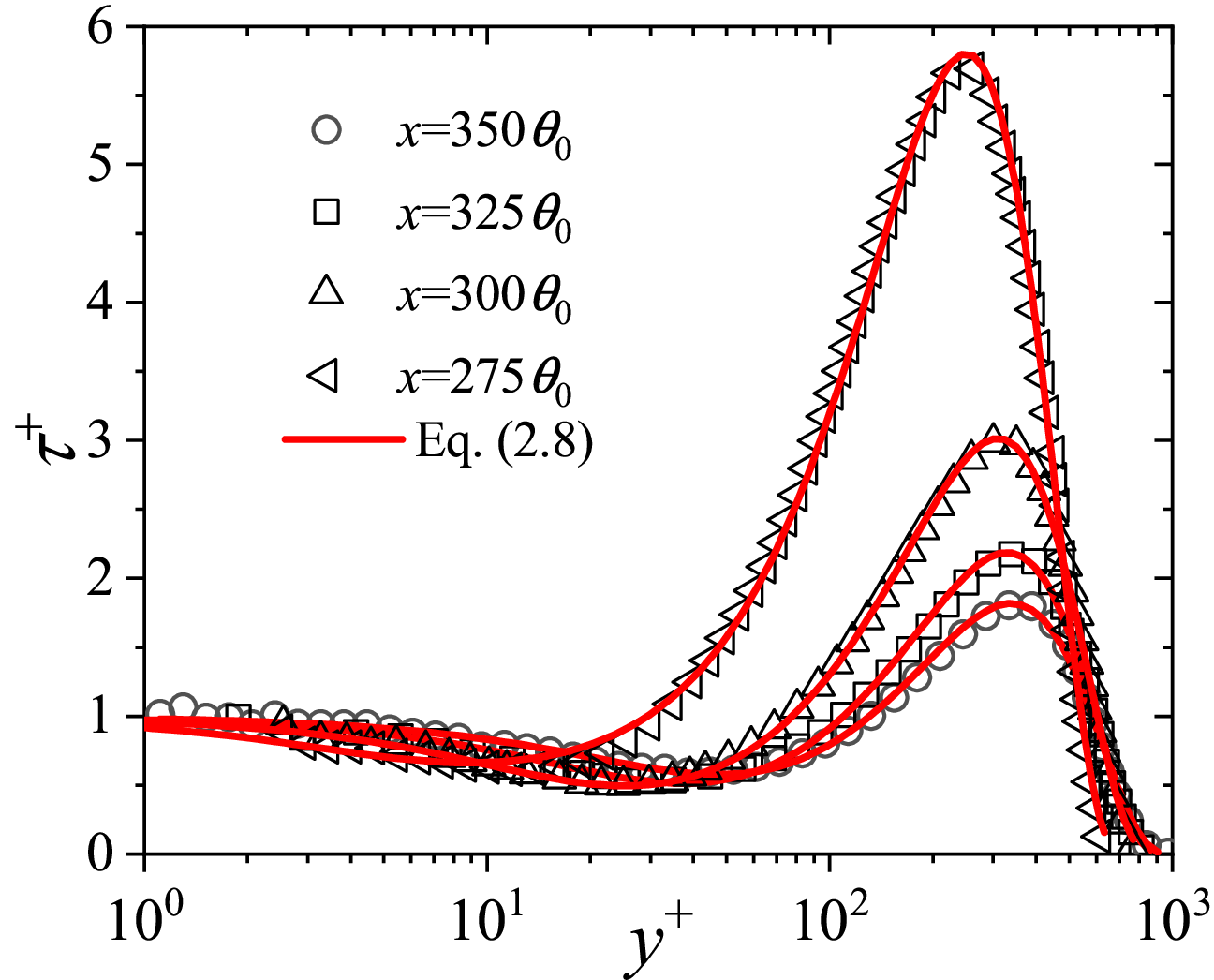}
    \includegraphics[width=0.5\linewidth]{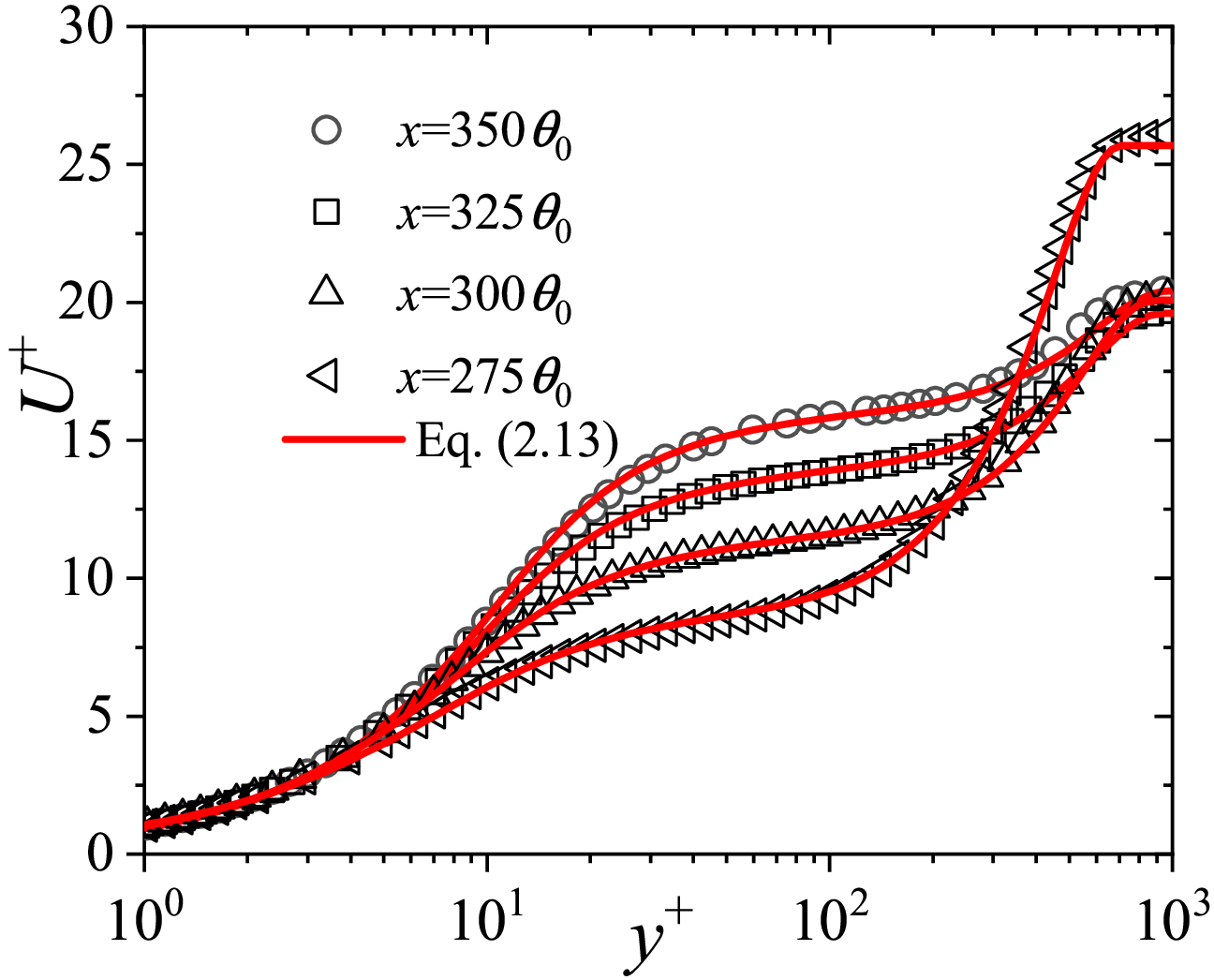}
  \caption{($a$) The total stress profiles, and ($b$) the mean velocity profiles of the APG TBLs in front of the separation bubble, ($c$) the total stress profiles, and ($d$) the mean velocity profiles of the reattached FPG TBLs after the separation bubble. Symbols denote the DNS data of \cite{abe2019direct}. Lines denote the current models. In ($b$), equation (\ref{eq:u_model}) is calculated with (\ref{eq:tau_PG}) and (\ref{eq:ell}). The dashed line indicates the viscous sublayer solution, and the dash-dot line indicates the overlap layer solution. The mean velocity profiles are vertically displaced for clarity. In ($d$), equation (\ref{eq:u_model}) is calculated with (\ref{eq:tau_FPG}) and (\ref{eq:ell_FPG}).}
  \label{fig:Abe_prediction}
\end{figure}

The current theory applies to the other streamwise locations in the Abe's DNS. The $\tau$ profiles are demonstrated in figures \ref{fig:Abe_prediction}($a$) and \ref{fig:Abe_prediction}($c$) for both the APG and FPG TBLs. The models describe the data at very high accuracy, independent of the flow development, which undergoes dramatic variations in the magnitude of the friction velocity and PG (the Clauser
pressure gradient parameter $\beta$ is 84 at $x=100\theta_0$, and -13.3 at $x=275\theta_0$).

The mean velocity profiles are predicted for the APG and FPG TBLs by the total stress and SL models, as shown in figures \ref{fig:Abe_prediction}($b$) and \ref{fig:Abe_prediction}($d$). In figure \ref{fig:Abe_prediction}($b$) the classical formulations for the viscous sublayer and overlap layer are also plotted for comparison with the current theory. The present prediction is more accurate (with uncertainty of about 1\%), and applies to the entire TBL. Figure \ref{fig:Abe_prediction} reveals that, the current models capture the prominent features of the APG and reattached FPG TBLs. For example, for the APG TBLs, the $U$ profile in the outer flow is significantly bent by a strong APG, leading to a intriguing velocity-deficit profile that has not been accurately predicted before, but captured by the current theory. For the reattached FPG TBL, a region with slowly increasing velocity occurs in the bulk of the boundary layer because of the intense turbulence coming from upstream separated shear layer, which is beyond the conventional approaches but has now been accurately described.

\begin{figure}
    \includegraphics[width=0.5\linewidth]{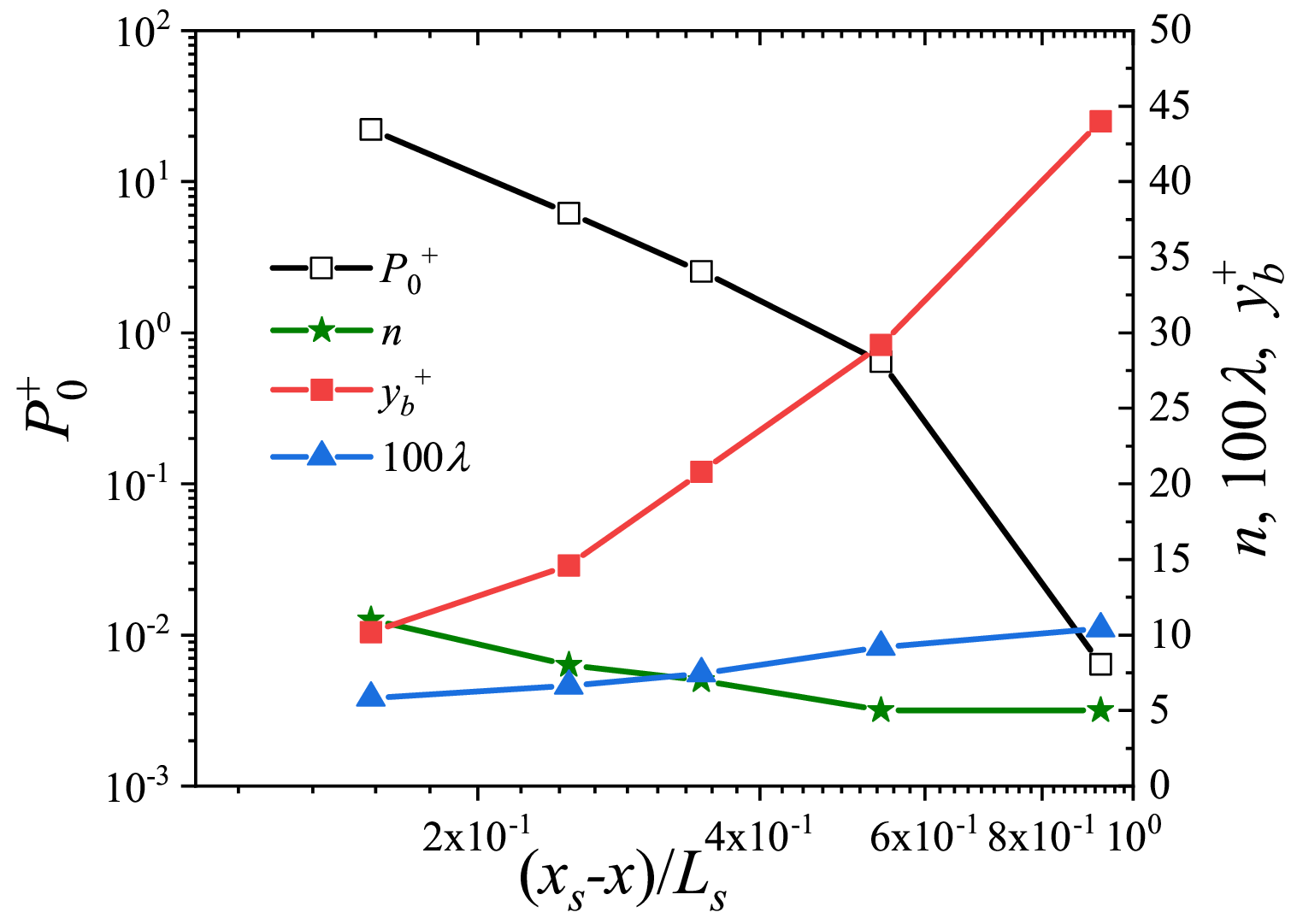}
    \includegraphics[width=0.5\linewidth]{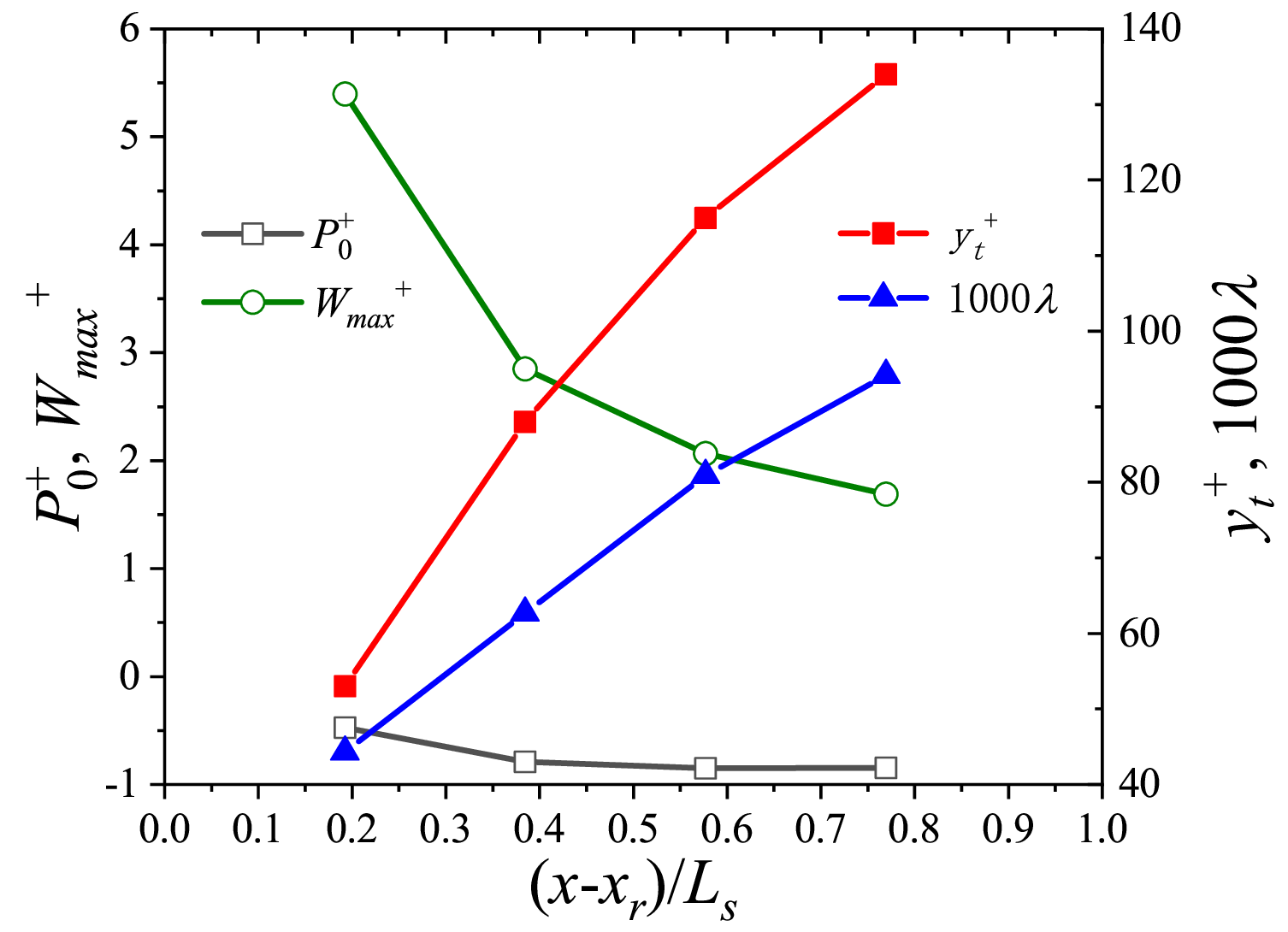}
  \caption{Streamwise variations of the MLS parameters in the total stress and SL models. ($a$) The APG TBL in front of the separation bubble, and ($b$) the reattached FPG TBL after the bubble. In ($a$), $\lambda=\kappa/n$, and in ($b$), $\lambda=\kappa/4$. $x_s$ denotes the separation point, $x_r$ denotes the reattachment point, and $L_s$ denotes the bubble width.}
  \label{fig:parameters_TBLs}
\end{figure}

There are only a minimum number of parameters in the current theory to characterize the complicated behaviors of the PG TBLs. Their variations with the streamwise coordinate are presented in figure \ref{fig:parameters_TBLs}. Note that these parameters are estimated separately. $P_0^+$ (or $P_0^+$ and $W_{max}^+$ for the reattached TBLs) is estimated by using a least squares method for the $\tau$ profile. $y_b^+$, $n$ and $\kappa$ (or $y_t^+$ and $\kappa$ for the reattached TBLs) are estimated by using a least squares method for the $U$ profile predicted with the $\tau$ and SL models.

For the APG TBL in front of the separation bubble (figure \ref{fig:parameters_TBLs}$a$), $P_0^+$ increases when $x$ approaches the separation point. Meanwhile, $y_b^+$ decreases and $n$ increases, which together lead to increasingly bent $U$ profile, as well as a wider outer flow --- the most apparent feature of the APG TBL. At the same time, $\lambda$ ($=\kappa/n$) decreases, meaning that the eddies near the upper boundary-layer edge become smaller comparing with the rapid increase of $\delta$.

For the reattached TBL (figure \ref{fig:parameters_TBLs}$b$), downstream the reattachment point $P_0^+$ is negative and decreasing, indicating a varying FPG. $W_{max}^+$ is decreasing also, showing the dissipation and relaxation of the intense turbulence from the separated shear layer. On the other hand, both $y_t^+$ and $\lambda$ increase with increasing $x$, revealing the growth of the eddy size and the dimensionless bulk-turbulent-layer thickness. The ever-increasing $y_t^+$ relaxes the action of the intense turbulence on the MLS of the TBL, such that a canonical MLS can be reconstructed when the PG eventually disappears.

\section{Summary}
In this paper a symmetry approach is proposed for modeling the entire mean velocity profiles of PG TBLs. First, a modified defect law is constructed for the total stress profile to model the PG and historical turbulence effects. Second, a multilayered power-law formulation is developed for the SL function following the structural ensemble dynamics theory. The SL model reveals that PG crucially affects the buffer layer thickness, the K\'{a}rm\'{a}n constant, and the outer-flow defect-law exponent. In case of intense turbulence from the upstream separated shear layer, a bulk turbulent layer replaces the conventional buffer layer and overlap layer. With the formulations the entire mean velocity profile is predicted analytically, and validated to accurately describe the published DNS data on 2D separation bubble.

The study presents a novel parameterization for the mean-flow profiles of general PG TBLs with a minimum number of empirical parameters, sound physical interpretations, and great accuracy, which paves a way to quantify the PG and historical turbulence effects in various boundary-layer flows. The success of the model reveals the MLS with the generalized dilation symmetry is the most important flow structure for characterizing complex TBLs.

\section*{Acknowledgements}
This research is supported by the National Natural Science Foundation of China under Grant No. 91952201.

\section*{References}

\bibliography{jfm}

\end{document}